\documentclass[12pt]{article}
\pdfoutput=1
\usepackage[T1]{fontenc}
\usepackage{graphicx,psfrag,epsf,color}
\usepackage{amsmath,amssymb,amsfonts}
\usepackage{physics}
\usepackage{siunitx}
\usepackage{array}
\usepackage{cite}
\usepackage{rotating}
\usepackage{slashed, cancel}
\usepackage{xcolor}
\usepackage{hyperref}
\usepackage[caption=false]{subfig}
\usepackage{graphicx}
\usepackage{mathrsfs}
\usepackage{float}
\restylefloat{table}

\usepackage{tabularx}
\usepackage{booktabs}
\usepackage{subfig}
\usepackage[margin=1.0in]{geometry}
\newcolumntype{C}{>{\centering\arraybackslash}X}
\usepackage{multirow} \usepackage{array}
\title{}
\author{andersrehult}
\date{}

\begin{document}

\begin{titlepage}

\vspace*{-2.0truecm}

\begin{flushright}
Nikhef-2023-002\\
SI-HEP-2023-05\\
P3H-23-014
\end{flushright}

\vspace*{1.3truecm}

\begin{center}
{
\Large \bf \boldmath New Perspectives for Testing Electron-Muon Universality}
\end{center}
\vspace{0.9truecm}

\begin{center}
{\bf Robert Fleischer\,${}^{a,b}$,  Eleftheria Malami\,${}^{a,c}$, Anders Rehult\,${}^{a}$, and  K. Keri Vos\,${}^{a,d}$}

\vspace{0.5truecm}

${}^a${\sl Nikhef, Science Park 105, NL-1098 XG Amsterdam,  Netherlands}

${}^b${\sl  Department of Physics and Astronomy, Vrije Universiteit Amsterdam,\\
NL-1081 HV Amsterdam, Netherlands}

${}^c${\sl  Center for Particle Physics Siegen (CPPS), Theoretische Physik 1,\\
Universität Siegen, D-57068 Siegen, Germany}

{\sl $^d$Gravitational 
Waves and Fundamental Physics (GWFP),\\ 
Maastricht University, Duboisdomein 30,\\ 
NL-6229 GT Maastricht, the
Netherlands}\\[0.3cm]

\end{center}

\vspace*{1.7cm}


\begin{abstract}
\noindent

Intriguing results for tests of the universality of electrons and muons through measurements of rates of $B\to K \ell^+ \ell^-$ and similar decays 
have been in the spotlight for years. The LHCb collaboration has recently reported new results which are in agreement with Lepton 
Flavour Universality, while the individual decay rates are found below their Standard Model predictions. In view of this new situation, we explore how 
much space is left for a violation of electron--muon universality. Considering new sources of CP violation and taking the new LHCb measurements 
into account, we show that significant differences between the short-distance coefficients for electronic and muonic 
final states are actually allowed by the current data. These patterns can be revealed through CP asymmetries in neutral and charged 
$B\to K \ell^+ \ell^-$ decays. We obtain correlations between these observables and map them to the short-distance coefficients. This results in regions in New Physics parameter space with large differences between CP asymmetries of the decays with final-state electrons and muons, thereby leaving a lot of room for possible surprises in the future high-precision era.
\end{abstract}


\vspace*{2.1truecm}

\vfill

\noindent
March 2023

\end{titlepage}

\thispagestyle{empty}

\vbox{}

\setcounter{page}{0}

\newpage
\section{Introduction}
A key feature of the Standard Model (SM) is the universality of electroweak couplings to different lepton flavours. This Lepton Flavour Universality (LFU) has been tested through experimental studies of $B$ meson decays, which in recent years have indicated that the universality may be violated \cite{BaBar:2012mrf,BELLE:2019xld,LHCb:2021trn}, leading to a lot of interest in the theoretical community (see for instance \cite{Alguero:2022wkd,Gubernari:2022hxn,Geng:2021nhg,Mahmoudi:2022hzx,SinghChundawat:2022zdf}). The evidence arises in two different decay classes (for a recent review, see \cite{Albrecht:2021tul}). First, in comparisons between decay rates of the charged-current  $B\to D^{(*)}\tau\bar\nu_\tau$ and $B\to D^{(*)}\mu\bar\nu_\mu$ channels, which originate at the tree level in the SM and whose differences are captured in the ratio $R_{D^{(*)}}$. Experimental measurements of this ratio disagree with theoretical predictions, raising the question of whether $\tau$--$\mu$ universality is violated. Second, which will be the focus of this paper, evidence arises in comparisons between the flavour-changing neutral current loop processes $B\to K^{(*)}\mu^+\mu^-$ and $B\to K^{(*)}e^+e^-$. The ratios $R_{K^{(*)}}$ of the corresponding decay rates are averaged over the CP-conjugate channels, although this is often not explicitly written. To make this difference explicit, we define  
\begin{equation}
   \langle R_K \rangle  \equiv \frac{ \Gamma(B^-\to K^-\mu^+\mu^-) +\Gamma(B^+\to K^+\mu^+\mu^-) }{\Gamma(B^-\to K^-e^+e^-)  + \Gamma(B^+\to K^+e^+e^-) } , 
    \label{eq:RKav}
\end{equation}
and equivalently for $R_{K^*}$. These LFU ratios are theoretically clean quantities, with the SM predicting values equal to 1 with excellent precision even when including tiny QED effects \cite{Bordone:2016gaq, Isidori:2022bzw}.

Before December 2022, the data indicated values of $R_{K^{(*)}}$ in the regime of
0.8, deviating from the SM value of 1 at the $3\,\sigma$ level and thereby suggesting a violation of the electron--muon universality of the SM. However, new results for $R_{K^{(*)}}$ were recently presented by the LHCb collaboration. In the $q^2 \in [1.1,6.0] \: \si{\giga\eV^2}$ bin of the momentum transfer 
$q^2$ to the $\ell^+\ell^-$ pair, the following new result was presented \cite{LHCb:2022qnv,LHCb:2022zom}: 
\begin{equation}\label{eq:rkmeas}
    \langle R_K \rangle = 0.949\pm 0.05 \ .
\end{equation}
A similar pattern is found for the $R_{K^*}$ ratio. This new result is consistent with the electron--muon universality of the SM. On the other hand, the experimental rates for $B\to K \mu^+\mu^-$ were not updated yet. Using the previous data set \cite{LHCb:2014cxe} and \eqref{eq:rkmeas}, also the $B\to K e^+e^-$ rates are now found to be below the SM predictions, with discrepancies up to the $4\,\sigma$ level with respect
to state-of-the-art calculations using the most recent input from lattice QCD calculations of the required hadronic form factors \cite{Parrott_SM_predictions,Parrott:2022rgu,Fleischer:2022klb}. Although these rates are not as clean as $\langle R_K \rangle$, as they depend on hadronic form factors, long-distance effects, and CKM matrix elements, they still hint towards New Physics (NP). 

At first sight, the new result for $\langle R_K \rangle$ in (\ref{eq:rkmeas}) may seem to imply that NP in $B \to~K\ell^+\ell^-$ should exhibit electron--muon universality and that deviations from it are largely constrained. However, as we will show in this paper this is actually not the case. The point is that new sources of CP violation, encoded in CP-violating phases of the short-distance Wilson coefficients, allow for LFU violation while being consistent with the new LHCb result. These effects would manifest themselves in different CP asymmetries for final states with electrons and muons, thereby providing an exciting way to test for electron--muon universality.

This paper is organized as follows: We first introduce the theoretical framework
and observables in Section~\ref{sec:theoframe}. In Section \ref{sec:tests}, we discuss the room for LFU violation in the current data, first considering only real NP couplings and then allowing the couplings to be complex. In Section~\ref{sec:Adire}, we present a new way to measure the direct CP asymmetry in the electron channel using LFU ratios. Finally, we conclude in Section~\ref{sec:con}.

\section{Theoretical Framework}
\label{sec:theoframe}
\subsection{Effective Hamiltonian}

The low-energy effective Hamiltonian for $b \to s \ell^+\ell^-$ transitions is \cite{Descotes-Genon:2020tnz, Altmannshofer:2008dz, Gratrex:2015hna,Buchalla:1995vs} 
\begin{equation}\label{eq:ham}
    \mathcal{H}_{\rm eff} = - \frac{4 G_F}{\sqrt{2}} \left[\lambda_u \Big\{C_1 (\mathcal{O}_1^c - \mathcal{O}_1^u) + C_2 (\mathcal{O}_2^c - \mathcal{O}_2^u)\Big\} + \lambda_t \sum\limits_{i \in I} C_i \mathcal{O}_i \right] \ ,
\end{equation}
where $\lambda_q = V_{qb} V_{qs}^*$ and $I = \{1c, 2c, 3, 4, 5, 6, 8, 7^{(\prime)}, 9^{(\prime)}\ell, 10^{(\prime)}\ell, S^{(\prime)}\ell, P^{(\prime)}\ell, T^{(\prime)}\ell\}$. We consider only light leptons, i.e., \ $\ell = \mu, e$, and neglect terms proportional to $\lambda_u\sim \mathcal{O}(5\%)$. We write the Wilson coefficients as
\begin{equation}\label{eq:Cwil}
    C_{i\ell} = C_i^{\rm SM} + C_{i\ell} ^{\rm NP} ,
\end{equation}
where in the SM \cite{Descotes-Genon:2013vna}
\begin{equation}\label{eq:C9sm}
    C_7^{\rm SM} = -0.292, \quad C_9^{\rm SM} = 4.07, \quad C_{10}^{\rm SM} = -4.31 \ ,
\end{equation}
are lepton flavour universal, and the $S,P$ and $T$ operators are absent. The latter operators would arise from NP which could also induce LFU violation couplings to operators, like $\mathcal{O}_9$ and $\mathcal{O}_{10}$, that are already present in the SM. 

The current-current operators $\mathcal{O}_1^c$ and $\mathcal{O}_2^c$ contribute via $c\bar{c}$ loops, which lead to long-distance effects through hadronic resonances (see, e.g., ~\cite{Khodjamirian:2012rm, Lyon:2013gba, Ciuchini:2015qxb, Capdevila:2017ert,Blake:2017fyh,Chrzaszcz:2018yza, Gubernari:2022hxn} for extensive discussions).  As in our previous work \cite{Fleischer:2022klb}, we take these effects into account through
\begin{equation}
    C_9^{\rm eff} = C_9 + Y(q^2) \ ,
\end{equation}
where $Y\equiv |Y| e^{i\delta_Y}$ with $\delta_Y$ defining a CP-conserving strong phase. We adopt the Kr\"uger--Sehgal parametrization \cite{Kruger:1996cv,Kruger:1996dt} and the fit by the LHCb collaboration \cite{LHCb:2016due} to their experimental data. We refer to these references and our previous work \cite{Fleischer:2022klb} for more details. This procedure resulted in four possible long-distance branches. In this work, we fix the long-distance model to the $Y_{--}$ branch and use the same model for the electron and muon channels. We emphasize that these choices do affect our results---nevertheless, it is useful to pick a specific long-distance model for our purpose of studying the possible differences between the muon and electron channels.

Keeping these points in mind, we expect that in the future the hadronic effects will be described with better precision. In fact, we showed that the different hadronic models leave distinct fingerprints on the CP asymmetries when considering different $q^2$ bins \cite{Fleischer:2022klb}. Therefore, improved measurements of the CP asymmetries in different $q^2$ bins could help narrow the theoretical uncertainty on the long-distance effects. 

In this paper, we consider NP contributions defined in \eqref{eq:Cwil} through 
\begin{equation}
    C_{i\ell}^{\rm NP} \equiv \abs{C_{i\ell}^{\rm NP}} e^{i \phi_{i\ell}^{\rm NP}} \ ,
\end{equation}
where $\phi_{i\ell}^{\rm NP}$ is a CP-violating NP phase. We only consider NP in $C_9$ and $C_{10}$, whose operators are defined as 
\begin{equation}
    \mathcal{O}_{9^{(\prime)}\ell} = \frac{e^2}{(4\pi)^2} [\bar s \gamma^\mu P_{L(R)} b] (\bar \ell \gamma_\mu \ell), \quad\quad \mathcal{O}_{10^{(\prime)}\ell} = \frac{e^2}{(4\pi)^2} [\bar s \gamma^\mu P_{L(R)} b] (\bar \ell \gamma_\mu \gamma_5 \ell),
\end{equation}
with $P_{R(L)} = \frac{1}{2} (1 \pm \gamma_5)$. We limit ourselves to these operators because they were often considered in global NP analyses. 

\subsection{Observables}
The key observable we consider in this paper is the CP-averaged LFU ratio $\langle R_K \rangle$ defined in \eqref{eq:RKav}. Even though the most recent LHCb measurement is now in agreement with the SM prediction of LFU, the individual $B\to K \mu^+\mu^-$ branching ratio still differs from the SM expectation. To clearly distinguish between the $B^-(\bar{B}^0_d)$ and $B^+(B^0_d)$ modes, we define the integrated decay rate of the $B^-$ as 
\begin{equation}
    \bar{\Gamma}_\ell[q^2_{\rm min}, q^2_{\rm max}] \equiv  \int_{q^2_{\rm min}}^{q^2_{\rm max}} dq^2 \frac{ d \Gamma(B^- \to K^- \ell^+\ell^-)}{dq^2}   
\end{equation}
and equivalently for $\bar{B}^0_d$. We denote by $\Gamma_\ell$ the equivalent expression for the $B^+\to K^+\ell^+\ell^-$ ($B^0_d$) mode. The angular distribution for $B^- \to K^- \ell^+\ell^-$ also gives interesting additional observables, but we do not consider these in the current work. Their expressions can be found, for instance, in \cite{Descotes-Genon:2020tnz, Gratrex:2015hna,Bobeth:2012vn}. 

The $q^2$-binned, CP-averaged LFU ratio is then
\begin{equation}
   \langle R_K \rangle [q_{\rm min}^2, q^2_{\rm max}]  \equiv \frac{\bar{\Gamma}_\mu[q_{\rm min}^2, q^2_{\rm max}] +\Gamma_\mu[q_{\rm min}^2, q^2_{\rm max}] }{\bar{\Gamma}_e[q_{\rm min}^2, q^2_{\rm max}] + \Gamma_e[q_{\rm min}^2, q^2_{\rm max}] } \ ,
    \label{eq:RKavfull}
\end{equation}
which in the SM is equal to 1 with excellent precision. Here, we consider $q_{\rm min}^2=1.1 \;{\rm GeV}^2$ and $q_{\rm max}^2 =6.0\; {\rm GeV}^2$. Compared to \eqref{eq:RKavfull}, the theoretical predictions for the individual branching ratios of $B\to K \mu^+\mu^-$ and $B\to K e^+ e^-$ have much larger uncertainties. First of all, the branching ratios depend on non-perturbative hadronic form factors, which despite having seen impressive progress in recent years \cite{Gubernari:2018wyi,Parrott:2022rgu} are still a key source of uncertainty. Second, the branching ratios depend on the hadronic long-distance effects discussed in the previous section, where the choice of long-distance model introduces a hidden systematic uncertainty. Finally, a large uncertainty comes from the CKM pre-factors. The branching ratios are directly proportional to $\lambda_t$ and by CKM unitarity to $|V_{cb}|$, for which there is a long-standing discrepancy between the exclusive and inclusive determinations \cite{Bordone_inclusive_Vbc, Bernlochner:2022ucr, Gambino:2020jvv} which leads to significantly different branching ratio predictions. These main sources of uncertainty cancel in the SM prediction of the LFU ratio in \eqref{eq:RKavfull}.

To determine the CKM elements, we use the inclusive/hybrid approach adapted in \cite{DeBruyn:2022zhw}, where the inclusive determination of $|V_{cb}|$ and the exclusive determination of $|V_{ub}|$ were used to perform a CKM fit. In our case, since we neglect higher order corrections entering via $\lambda_u$, the inclusive and hybrid scenario coincide. From \cite{DeBruyn:2022zhw}, we have $\abs{\lambda_t} = \abs{V_{tb}V_{ts}^*}=(41.4~\pm~0.5)~\times~10^{-3} $. Using this input, we recently calculated the branching ratio in the $1.1 \;{\rm GeV}^2 <q^2 < 6.0 \; {\rm GeV}^2$ range \cite{Fleischer:2022klb} to be
\begin{equation}
    \mathcal{B}(B^\pm \to K^\pm \mu^+\mu^-)^{\rm SM}[1.1,6.0]_{\rm incl/hybrid} = (1.83 \pm 0.14) \times 10^{-7} \ ,
    \label{eq:SM_BR_lowq2}
\end{equation}
where we used the most recent lattice calculation for the form factors \cite{Parrott:2022rgu} and we added the variation of the different long distance branches as an additional uncertainty.
Interestingly, the measurement of the CP-averaged $B^\pm \to K^\pm \mu^+\mu^-$ branching ratio from the LHCb collaboration\cite{LHCb:2014cxe},
\begin{equation}
    \mathcal{B}(B^\pm \to K^\pm \mu^+\mu^-) = (1.19 \pm 0.07)  \times 10^{-7} \ ,
    \label{eq:brexp}
\end{equation}
is lower than our prediction, hinting at NP with a statistical significance of $3.5~\sigma$.

We can access important information also through the direct and mixing-induced CP asymmetries. The charged $B^\pm$ decays exhibit only a direct CP asymmetry, defined through 
\begin{equation}
    \mathcal{A}_{\rm CP}^{\rm dir}[q^2_{\rm min}, q^2_{\rm max}] = \frac{\bar \Gamma[q^2_{\rm min}, q^2_{\rm max}] - \Gamma[q^2_{\rm min}, q^2_{\rm max}]}{\bar \Gamma[q^2_{\rm min}, q^2_{\rm max}] + \Gamma[q^2_{\rm min}, q^2_{\rm max}]} \ .
    \label{eq:q2_binned_CP_asymm}
\end{equation}
 
For neutral $B_d^0$ decays, $B_d^0$--$\bar{B}_d^0$ oscillations give rise to a mixing-induced CP asymmetry through interference between the $B_d^0$ and $\bar{B}_d^0$ mesons decaying into the same final state. The $B_d^0 \to K^0 \ell^+\ell^-$ and $\bar{B}_d^0 \to \bar{K}^0 \ell^+\ell^-$ are flavour specific decays, for which the subsequent $K^0 (\bar{K}^0)$ decay determines the initial $B_d^0$ flavour. Observing the neutral kaons as a $K_{\rm S}$ final state, on the other hand, does allow both $B^0_d$ and $\bar{B}_d^0$ to decay into the same final state, thereby generating a mixing-induced CP asymmetry. This observable is accessible through the time-dependent rate. Following the definition in \cite{Fleischer:2022klb}, we define 
\begin{equation}
\begin{aligned}
    \frac{\Gamma(\bar B^0_d(t) \to K_{\rm S} \ell^+\ell^-) 
    - \Gamma(B^0_d(t) \to K_{\rm S} \ell^+\ell^-)}{\Gamma(\bar B^0_d(t) \to K_{\rm S} \ell^+\ell^-)
    + \Gamma(B^0_d(t) \to K_{\rm S} \ell^+\ell^-)} 
    &=\frac{\mathcal{A}_{\rm CP}^{\rm dir,\ell}\cos(\Delta M_d t)  + \mathcal{A}^{\rm mix,\ell}_{\rm CP} \sin (\Delta M_d t) }{\cosh (\frac{\Delta \Gamma_{d}}{2} t) + \mathcal{A}_{\Delta \Gamma}^\ell \sinh (\frac{\Delta \Gamma_{d}}{2} t)} \ ,
\end{aligned}
\label{eq:time_dependent_CP_asymmetry}
\end{equation}
where $\Delta\Gamma_d \equiv \Gamma^L_d - \Gamma^H_d$ is the decay-width difference in the $B_d$ system, which is negligible, and $\Delta M_d = M^{H}_{d} - M^L_d$ is the mass difference between the ``heavy'' and ``light'' eigenstates. Here we neglect CP violation in the kaon system and treat the $K_{\rm S}$ as a CP eigenstate. Due to isospin symmetry between the spectator quarks, the direct CP asymmetry of the charged $B^\pm$ decay and neutral $B_d^0$ decays are equal. The expressions for $\mathcal{A}_{\rm CP}^{\rm dir,\ell}$ and $\mathcal{A}_{\rm CP}^{\rm mix,\ell}$ in terms of the form factors and Wilson coefficients can be found in \cite{Fleischer:2022klb, Descotes-Genon:2015hea}.

In the SM we have 
\begin{equation}\label{eq:smpred}
\mathcal{A}_{\rm CP}^{\rm dir}|(B_d^0\to K_{\rm S} \ell^+\ell^-)|_{\rm SM} = 0, \quad\quad   \mathcal{A}_{\rm CP}^{\rm mix}(B_d^0\to K_{\rm S} \ell^+\ell^-)|_{\rm SM}= 0.72 \pm 0.02 \ , \end{equation}
where we used for the $B_d^0$--$\bar{B}_d^0$ mixing phase $\phi_d= (44.4\pm 1.6)^\circ$\cite{Barel:2020jvf,Barel:2022wfr}. The direct CP asymmetry is zero because we neglect the tiny CKM-suppressed $\lambda_u$ terms. As we showed in \cite{Fleischer:2022klb}, the direct CP asymmetry is rather sensitive to the specific choice of hadronic long-distance model, which can be turned into a positive property using the fingerprinting strategy discussed in that reference. On the other hand, the mixing-induced CP asymmetry is much more robust with respect to long-distance model, which was already pointed out in \cite{Descotes-Genon:2020tnz}.

\section{Testing Electron-Muon Universality}\label{sec:tests}
The previous measurements indicating $\langle R_K \rangle< 1$ inspired a plethora of NP explanations with LFU-violating couplings to electrons and muons \cite{Buttazzo:2017ixm,Bordone:2019uzc,Bordone:2017bld,Bordone:2018nbg, Cornella:2021sby,Smolkovic:2022axy,Ciuchini:2022wbq}. 
In the following, we consider the recent measurement of $R_K$ in \eqref{eq:rkmeas} and explore what room remains for LFU-violating couplings. We first consider real couplings before going to the general case of complex NP couplings.

\subsection{CP-conserving New Physics Contributions}
\begin{figure}[t!]
    \centering
    \subfloat{\includegraphics[width=0.3\textwidth]{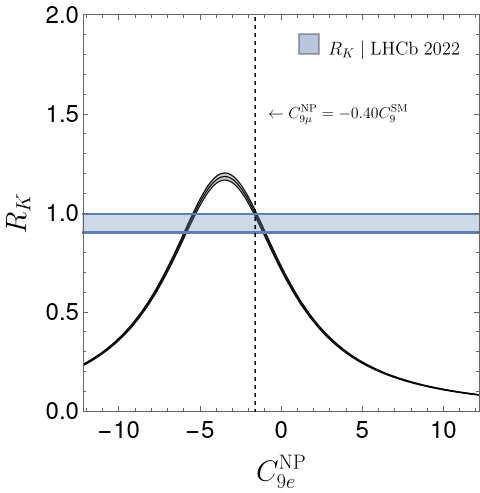}}
    \hfill
    \subfloat{\includegraphics[width=0.3\textwidth]{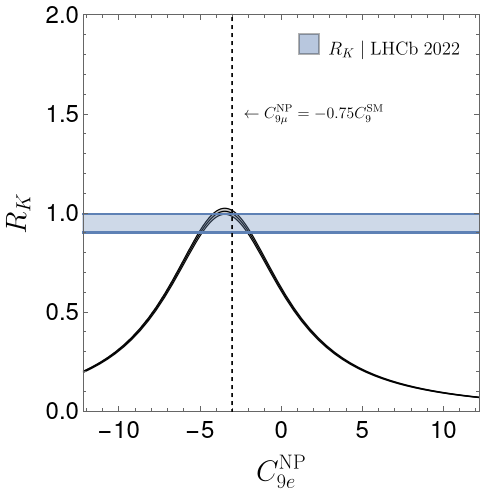}}
    \hfill
    \subfloat{\includegraphics[width=0.3\textwidth]{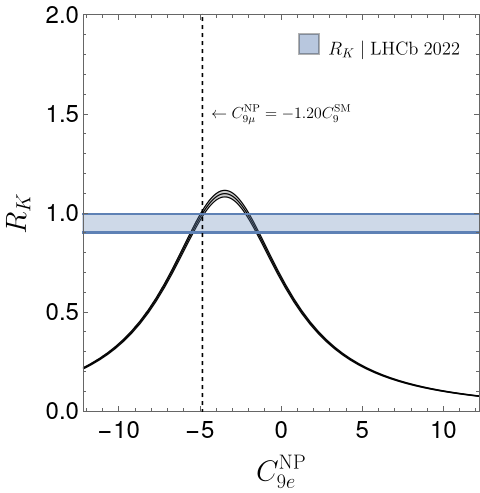}}
    \caption{The $\langle R_K \rangle$ ratio as a function of $C_{9e}^{\rm NP}$ for Scenario 1. Each of the plots corresponds to a different benchmark point for $C_{9\mu}^{\rm NP}$.}
    \label{fig:RKrealC9e_2}
\end{figure}

\begin{figure}[t]
    \centering
    \subfloat{\includegraphics[width=0.3\textwidth]{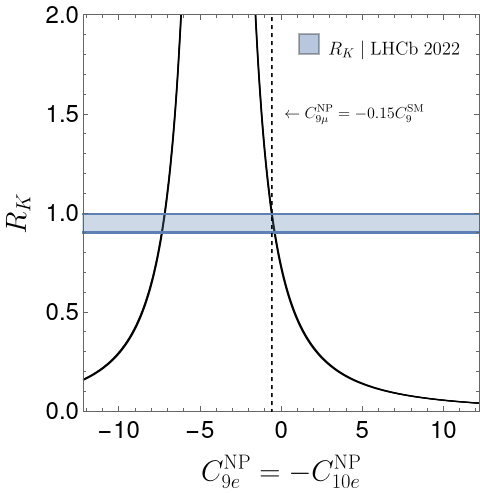}}
    \hfill
    \subfloat{\includegraphics[width=0.3\textwidth]{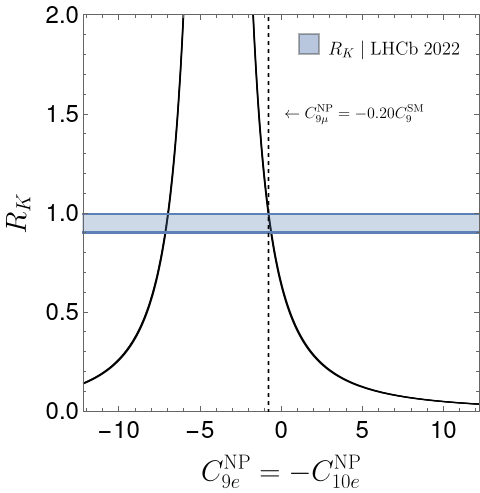}}
    \hfill
    \subfloat{\includegraphics[width=0.3\textwidth]{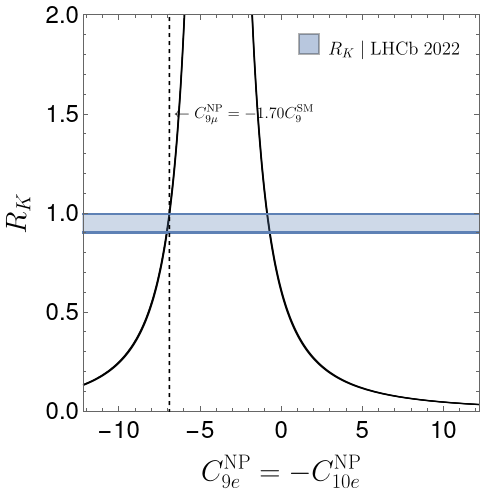}}
    \caption{The $\langle R_K \rangle$ ratio as a function of $C_{9e}^{\rm NP}=-C_{10e}^{\rm NP}$ for Scenario 2. Each of the plots corresponds to a different benchmark point for the muon channel with $C_{9\mu}^{\rm NP}=-C_{10\mu}^{\rm NP}$.  }
    \label{fig:RKrealC9e_3}
\end{figure}
We will consider two different scenarios: 
\begin{align}
    {\rm Scenario~1:} & \quad C_{9\mu}^{\rm NP}\quad {\rm only} \ , \\
 {\rm Scenario~2:} &  \quad C_{9\mu}^{\rm NP} = -C_{10\mu}^{\rm NP} \ ,
\end{align}
which we also discussed in our previous work \cite{Fleischer:2022klb} and are often discussed in the literature. 

The discrepancy between the theoretical prediction and the experimental measurement of the $B\to K \mu\mu$ branching ratio in \eqref{eq:SM_BR_lowq2} and \eqref{eq:brexp}, respectively, can be reduced with negative values of $C_{9\mu}^{\rm NP}$ and/or positive values of $C_{10\mu}^{\rm NP}$. NP contributions with these signs will lower the branching ratio from the SM value down towards the experimental measurement. Reducing the tension to $1\sigma$ and lower, we find the following ranges for $C_{9\mu}^{\rm NP} (= -C_{10\mu}^{\rm NP})$:
\begin{equation}
\begin{aligned}\label{eq:range}
  { \rm Scenario~1:} \quad C_{9\mu}^{\rm NP} &= [-1.32, -0.40]C_9^{\rm SM},\\
  { \rm Scenario~2:}   \quad C_{9\mu}^{\rm NP} &= -C_{10\mu}^{\rm NP}  = [-0.23, -0.15]C_9^{\rm SM} \; \cup \; [-1.76, -1.69]C_9^{\rm SM}\ ,
\end{aligned}
\end{equation}
where $C_9^{\rm SM}$ is given in \ref{eq:C9sm}. These ranges can also be read off from Fig.~\ref{fig:morebenchmarkpoints}, which we discuss in the next Section, at $\phi=180^\circ$.  

To show the allowed range for $C_{9e}$ given by $\langle R_K \rangle$ in \eqref{eq:rkmeas}, we fix $C_{9\mu}^{\rm NP}$ to three different values that fall within the ranges in \eqref{eq:range}. For Scenario~1, we show in Fig.~\ref{fig:RKrealC9e_2} the corresponding value of $\langle R_K \rangle$ as a function of $C_{9e}^{\rm NP}$ for these three fixed values, together with the current experimental $1\sigma$ range for $\langle R_K \rangle$ from \eqref{eq:rkmeas}. We include theoretical uncertainties from the form factors and the variation of long-distance branches. We observe that there is always a solution that is consistent with LFU (dotted line) within uncertainties because the $\langle R_K \rangle$ measurement is just slightly more than $1\sigma$ lower than the SM. On top of that, there is a second solution for $C_{9e}^{\rm NP}$ that is (significantly) different from the muon coefficient. In the middle panel of Fig.~\ref{fig:RKrealC9e_2} these solutions are rather close such that the allowed region for $C_{9e}^{\rm NP}$ spans a range of values that also includes the LFU solution. The constraints for Scenario~2 given in Fig.~\ref{fig:RKrealC9e_3} show the same behaviour, although in this second scenario there is less room left for LFU-violating real NP Wilson coefficients. Finally, we note that for real NP couplings, both the CP asymmetries in $B \to K\ell^+\ell^-$ are unaffected and take their SM values, given in~\eqref{eq:smpred}. 

\subsection{CP-violating New Physics Contributions}
We now allow for complex NP couplings with new sources of CP violation in both the electron and muon sectors and explore how much room there is for LFU-violating NP in such a scenario. To do so, we proceed as follows:
\begin{enumerate}
    \item Constrain the muonic Wilson coefficients by using the experimental branching ratio and direct CP asymmetry of $B^-\to K^- \mu^+\mu^-$.
    \item Combine the constraints on muonic coefficients with the new $\langle R_K \rangle $ measurement in \eqref{eq:rkmeas} to constrain the electronic Wilson coefficients.
    \item Compute the direct and mixing-induced CP asymmetries of $B^0_d \to K_{\rm S} e^+ e^-$ that could arise in the electron sector within the constraints. 
    \end{enumerate}
\begin{figure}
    \centering
\includegraphics[width=1\textwidth]{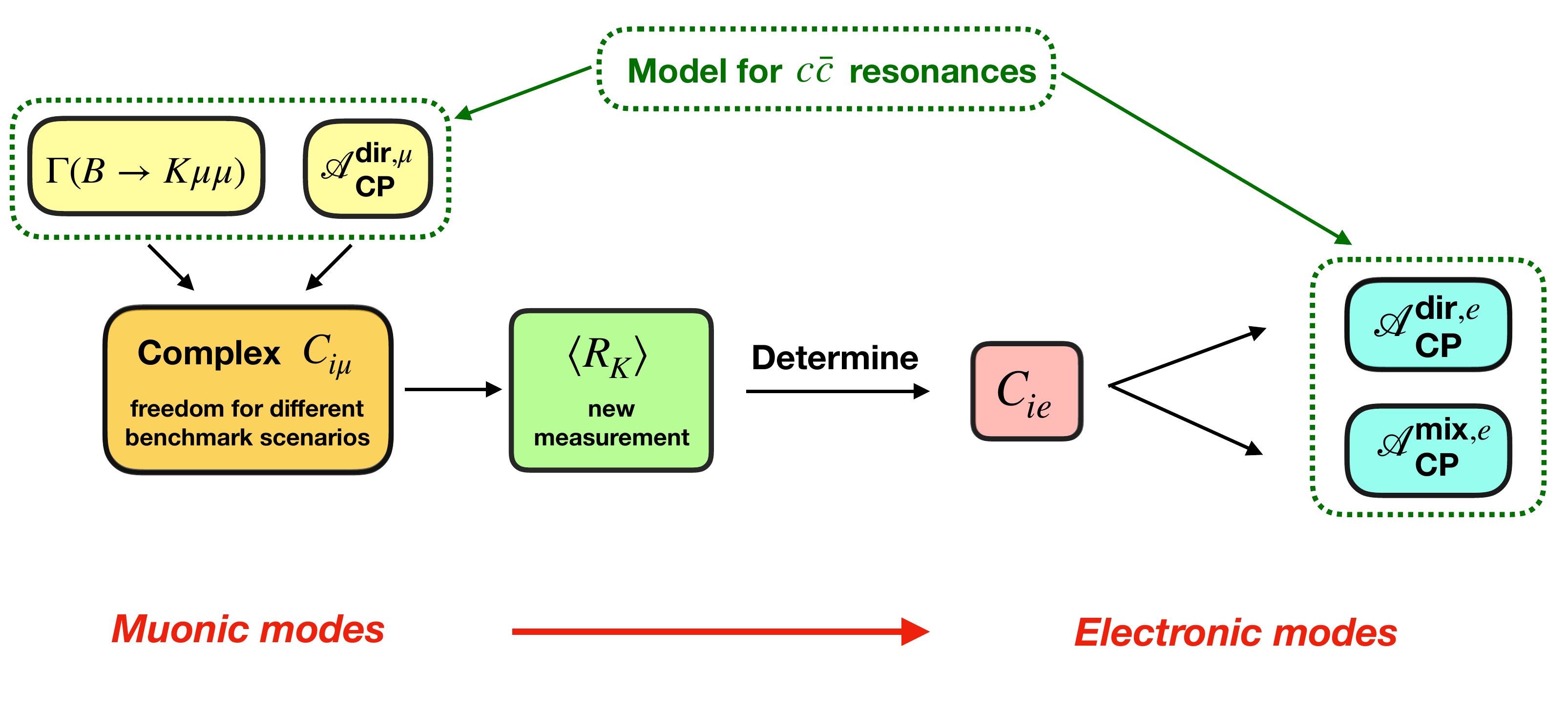}
\caption{Illustration of our procedure to determine the NP signals in the electronic CP asymmetries.}
\label{fig:flow}
\end{figure}
We schematically illustrate this procedure in Fig.~\ref{fig:flow}.

\begin{figure}[t!]
    \centering
    \subfloat{
    \includegraphics[width=0.4\textwidth]{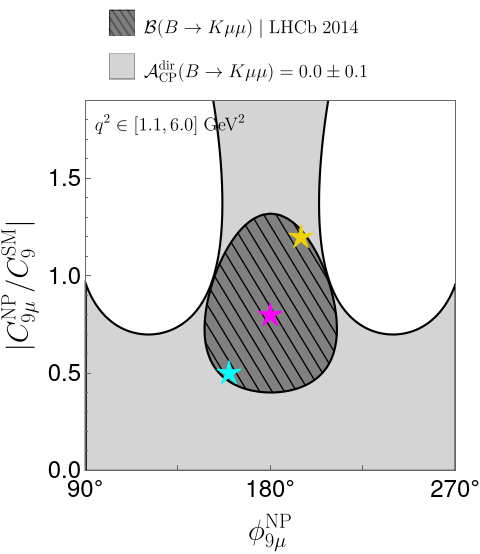}}
    \hfill
     \subfloat{
    \includegraphics[width=0.4\textwidth]{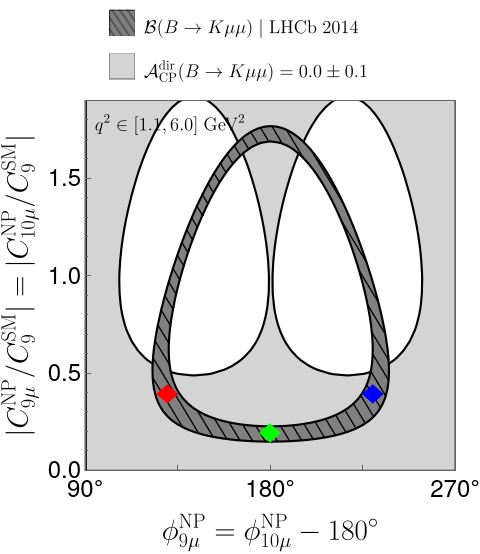}}
    \caption{Current bounds on $C_{9\mu}^{\rm NP}$ for NP Scenario 1 (left) and Scenario 2 (right) from the experimental measurements of the $B\to K\mu^+\mu^-$ branching ratio (gray) and the direct CP asymmetry (striped). The stars and diamonds indicate benchmark points as discussed in the text.} 
    \label{fig:morebenchmarkpoints}
\end{figure}

To obtain the current bounds on the muon coefficients, we use constraints on the branching ratio and direct CP asymmetry of $B^- \to K^-\mu^+\mu^-$. For the branching ratio, we use the measurement in \eqref{eq:brexp}, which we stress is lower than
the SM prediction. Regarding the direct CP asymmetry, this observable has been constrained by the LHCb collaboration in different $q^2$ bins \cite{LHCb:2014mit}. An average over the $q^2$ bins gives a rather small CP asymmetry of $\mathcal{A}_{\rm CP}^{\rm dir}(B\to K\mu^+\mu^-) = 0.012 \pm 0.017$. However, taking this average assumes that the asymmetry is constant across the spectrum. This is not the case, as the asymmetry will be enhanced near $c \bar c$ resonance peaks \cite{Becirevic:2020ssj}, regions that were vetoed in the LHCb study. For this reason, we use a more conservative constraint 
\begin{equation}
    \mathcal{A}_{\rm CP}^{\rm dir}(B^-\to K^-\mu^+\mu^-) = 0.0 \pm 0.1 \ .
\end{equation}
Using these values, we obtain the allowed parameter space for $C_{9\mu}$ in Fig. \ref{fig:morebenchmarkpoints}. The left figure shows that the measured value of the branching ratio indicates that $\abs{C_{9\mu}^{\rm NP}}$ deviates from 0, as already discussed, specifically ranging from 40 to 130$\%$ of $\abs{C_{9\mu}^{\rm SM}}$. For $C_{9\mu}^{\rm NP} = -C_{10\mu}^{\rm NP}$ in Fig.~\ref{fig:morebenchmarkpoints} (right) a value between 15 and 170$\%$ of $\abs{C_{9\mu}^{\rm SM}}$ is required. Within the experimental measurements, there is considerable freedom in the magnitude and phase of $C_{9\mu}^{\rm NP} (= -C_{10\mu}^{\rm NP}$). For convenience, we define six benchmark points that cover the allowed region listed in Table~\ref{tab:benchmark}. 

\begin{table}[t!]
    \centering
    \begin{tabular}{lc|lc}\toprule
    \multicolumn{2}{c}{$C_{9}^{\rm only}$ (Stars)}  & \multicolumn{2}{c}{$C_9^{\rm NP} = -C_{10}^{\rm NP}$ (Diamonds)}\\ \hline
Cyan & $0.50 \abs{C_9^{\rm SM}} e^{i 160^\circ}$ & Red & $0.40\abs{C_9^{\rm SM}} e^{i 130^\circ}$  \\
Magenta & $0.80 \abs{C_9^{\rm SM}} e^{i 180^\circ}$  & Green & $0.20 \abs{C_9^{\rm SM}} e^{i 180^\circ}$ \\
Yellow & $1.20 \abs{C_9^{\rm SM}} e^{i 195^\circ}$ &  Blue & $0.40 \abs{C_9^{\rm SM}} e^{i 230^\circ}$ \\
  \bottomrule
    \end{tabular}
    \caption{Six different benchmark points that cover the allowed regions for $C_{9\mu}$.}
    \label{tab:benchmark}
\end{table}

We can now use these constraints and benchmark points to show what the $\langle R_K \rangle$ measurement implies for NP in the electron sector. Since complex NP couplings will leave distinct signals in the CP asymmetry space, we consider here neutral $B_d^0\to K_{\rm S} \ell^+ \ell^-$ decays, which give us access to both direct and mixing-induced CP asymmetries. We note that the direct CP asymmetry for this neutral mode and the charged mode $B^- \to K^- \ell^+ \ell^-$ are equal as we do not take isospin corrections into account and consider the same hadronic long-distance model for both modes.

\begin{figure}[t!]
    \centering
        \subfloat{\includegraphics[width=0.33\textwidth]{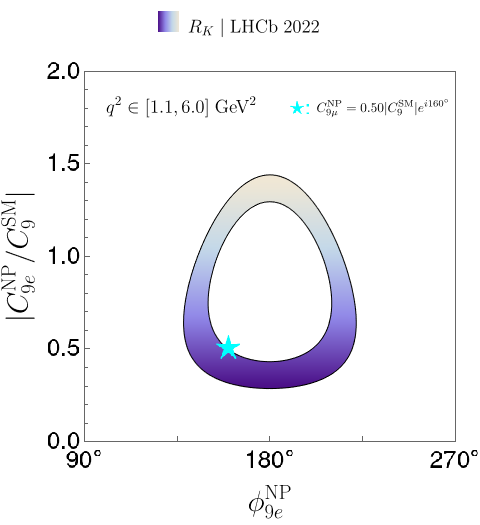}}
 \subfloat{
    \includegraphics[width=0.33\textwidth]{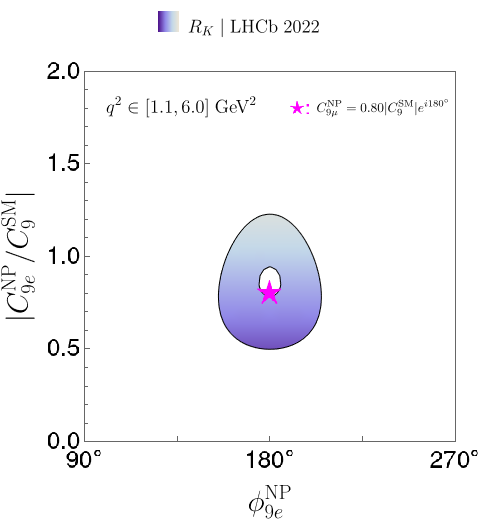}}
     \subfloat{
    \includegraphics[width=0.33\textwidth]{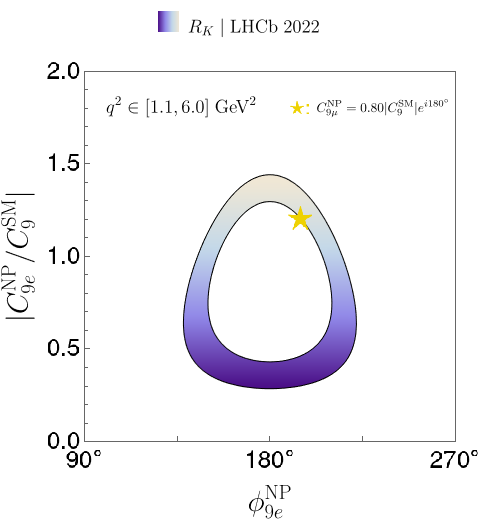}}\\
     \subfloat{
    \includegraphics[width=0.33\textwidth]{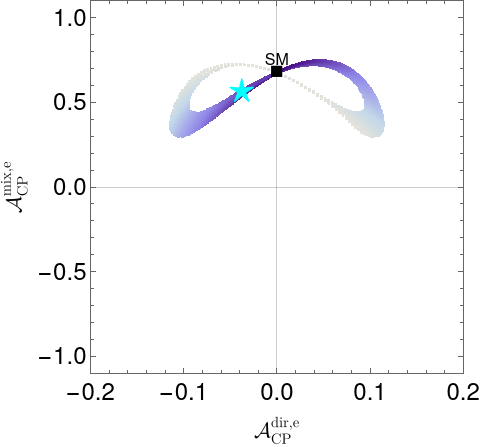}}
    \subfloat{
    \includegraphics[width=0.33\textwidth]{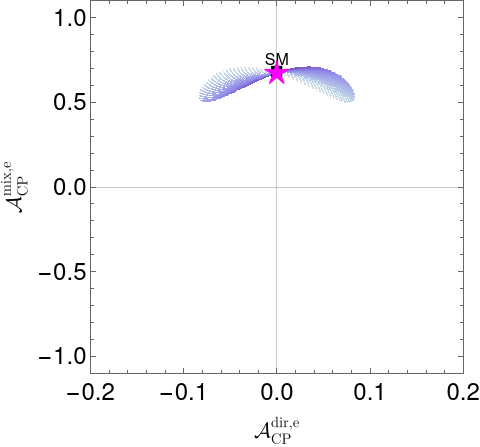}}
     \subfloat{
    \includegraphics[width=0.33\textwidth]{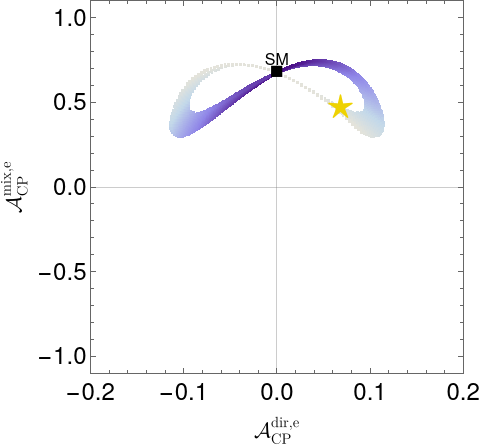}} 
    \caption{Constraints on $C_{9e}^{\rm NP}$ (upper) for the different benchmark points of Scenario 1 and the corresponding constraints on the CP asymmetries (lower).}
\label{fig:plots_star}
\end{figure}

In Fig.~\ref{fig:plots_star}, we show the allowed regions for $C_{9e}^{\rm NP}$ and its phase $\phi_{9e}^{\rm NP}$ for the star benchmark points belonging to Scenario~1, together with the corresponding direct and mixing-induced CP asymmetries. The colour coding shows the one-to-one correspondence between the value of $C_{9e}^{\rm NP}$ and the CP asymmetries. We find that the electronic Wilson coefficient can have a magnitude and phase strikingly different from the muonic one. The magnitude ranges from slightly smaller to more than twice as big. In other words, even with a value of $\langle R_K \rangle$ close to unity, there could be significant LFU violation hiding in the data. Importantly, we also observe that the electronic CP asymmetries can assume any values along the curve in the figure, resulting in CP asymmetries that are significantly different from those in the muon channel (indicated by the star). For completeness, we also mark the SM points. 

\begin{figure}[t!]
    \centering
    \subfloat{
    \includegraphics[width=0.33\textwidth]{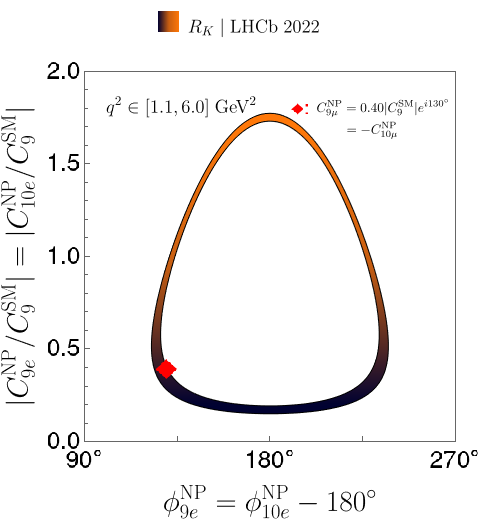}}
     \subfloat{
    \includegraphics[width=0.33\textwidth]{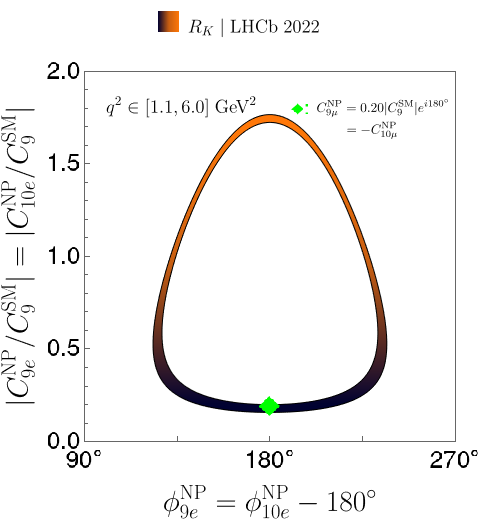}}
     \subfloat{
    \includegraphics[width=0.33\textwidth]{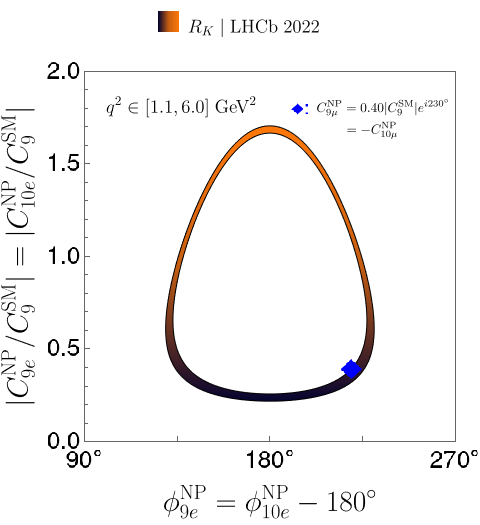}}\\
     \subfloat{
    \includegraphics[width=0.33\textwidth]{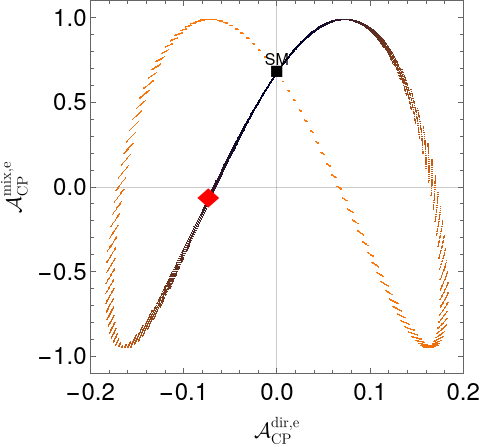}}
        \subfloat{
    \includegraphics[width=0.33\textwidth]{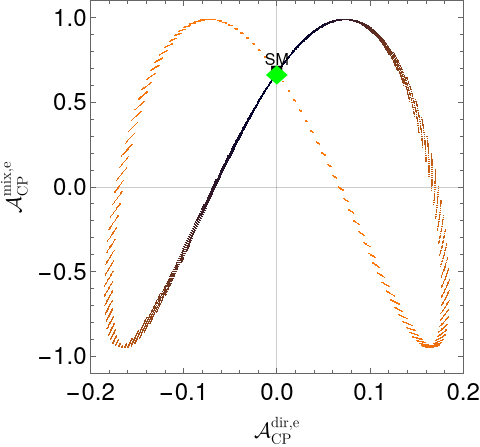}}
        \subfloat{
    \includegraphics[width=0.33\textwidth]{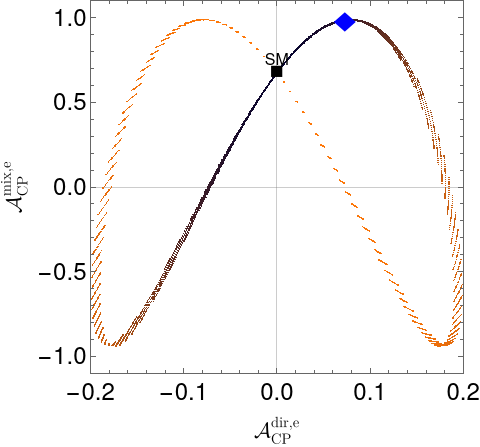}}
    \caption{Constraints on $C_{9e}^{\rm NP}$ (upper) for the different benchmark points of Scenario 2 and the corresponding constraints on the CP asymmetries (lower).}
\label{fig:plots_diamond} 
\end{figure}

We stress that this conclusion is independent of both the choice of NP scenario and the specific benchmark value for $C_{9\mu}^{\rm NP}$. To illustrate this, we show in Fig.~\ref{fig:plots_diamond} our results for the diamond benchmark points belonging to Scenario~2. While the exact shape and size of the allowed regions for $C_{9e}^{\rm NP}$ vary with the choice of $C_{9\mu}^{\rm NP}$, the fact remains that the muonic and electronic Wilson coefficients can be significantly different.

Our analysis demonstrates that even with a value of $\langle R_K \rangle$ close to unity, the NP Wilson coefficients that enter $B \to Ke^+e^-$ can strongly differ from the NP Wilson coefficients of $B \to K\mu^+\mu^-$. Beside the magnitude of the Wilson coefficients, also their complex, CP-violating phases can differ significantly. The latter difference implies that the electron and muon channels can have distinctly different CP asymmetries, a key point for experimentally testing LFU violation.

Clearly, to find out whether NP in $B\to K\mu^+\mu^-$ could differ from the effects in $B\to~K~e^+~e^-$, precise measurements of the CP asymmetries in both channels are required. So far, the only limit on the electronic mode comes from the Belle Collaboration\cite{Belle:2009zue}:
\begin{equation}\label{eq:Adir_e}
    \mathcal{A}_{\rm CP}^{\rm dir, e} = 0.14 \pm 0.14 \ , 
\end{equation}
which is a weighted average over different $q^2$ bins, both below and above the $c \bar c$ resonances. Within $(1-2)~\sigma$ all our results lie in this range. With future measurements in both the electron and muonic channel for different $q^2$ bins, we can fully utilise our fingerprinting method presented in \cite{Fleischer:2022klb} to determine the Wilson coefficients separately and unambiguously for the muonic and the electronic measurements. Given the significant room for new CP-violating couplings which violate electron--muon universality, we strongly encourage the experimental community to make detailed feasibility studies and perform the corresponding measurements.

\section{\boldmath{$R_K$} \unboldmath Ratios as Probes of Direct CP Violation}\label{sec:Adire}
Our previous studies show that there is quite some room for new CP-violating couplings to violate electron-muon universality. In this case, there is also a difference between the LFU ratios in the $B^-$ and $B^+$ modes, (and equivalently between $B_d^0$ and $\bar{B}_d^0$). Specifically, we can define
\begin{equation}
 R_K \equiv \frac{\Gamma_\mu}{\Gamma_e} \ , \quad\quad\quad  \bar{R}_K \equiv \frac{\bar{\Gamma}_\mu}{\bar{\Gamma}_e} \ ,
    \label{eq:RK}
\end{equation}
for the $B^+$ and its CP-conjugate $B^-$ mode, respectively. In this case, the averaged quantity $ \langle R_K \rangle$ in \eqref{eq:RKav} is not the same as the sum of $R_K$ and $\bar{R}_K$ as defined in \eqref{eq:RK}. 
Therefore, we highlight that it is very interesting to measure both ratios separately. In fact, the CP-averaged LFU ratio can be written as \cite{Fleischer:2022klb}
\begin{equation}
    \langle R_K \rangle = \frac{1}{2} \left[R_K + \bar{R}_K + (\bar{R}_K - R_K) \mathcal{A}_{\rm CP}^{\rm dir,e}\right] \ ,
\end{equation}
where the direct CP asymmetry for the electron mode enters. 
We can then also express the amount of CP violation in $R_K$ by defining a new observable \cite{Fleischer:2022klb}
\begin{equation}
\mathcal{A}_{\rm CP}^{R_K} \equiv \frac{\bar R_K - R_K}{\bar R_K + R_K} \ ,
\label{eq:CPV_in_RK}
\end{equation}
which is currently not measured. This new observable in \eqref{eq:CPV_in_RK} provides a measure of whether lepton flavour non-universal NP in $B \to K\ell^+\ell^-$ is also CP violating. We can rewrite \eqref{eq:CPV_in_RK} as
\begin{equation}
\begin{aligned}
\mathcal{A}_{\rm CP}^{R_K} &=  \left[ \frac{\mathcal{A}_{\rm CP}^{\rm dir,\mu} - \mathcal{A}_{\rm CP}^{\rm dir,e}}{1 - \mathcal{A}_{\rm CP}^{\rm dir,\mu}\ \mathcal{A}_{\rm CP}^{\rm dir,e}} \right] \ .
\label{eq:CP_separated_RK_direct_CP_asymmetries}
\end{aligned}
\end{equation}
Interestingly, measuring the LFU ratios for the CP-conjugated modes separately also gives a new way to determine the electron direct CP asymmetry \cite{Fleischer:2022klb} 
\begin{equation}\label{eq:ACPRK_Adire}
\begin{aligned}
    \mathcal{A}_{\rm CP}^{\rm dir,e} &= \frac{2 \langle R_K \rangle - R_K - \bar R_K}{\bar R_K - R_K}\\&= \frac{2 \langle R_K \rangle}{\bar{R}_K - R_K} - \frac{1}{\mathcal{A}_{\rm CP}^{R_K}} \ .
\end{aligned}
\end{equation}
Alternatively, we can also access the direct CP asymmetry through: 
\begin{equation}
    \mathcal{A}_{\rm CP}^{\rm dir, e} = \frac{\mathcal{A}_{\rm CP}^{R_K}}{\mathcal{A}_{\rm CP}^{\rm dir, \mu}\left( 1 + \mathcal{A}_{\rm CP}^{R_K} \right)} \ .
\end{equation}
Considering the large uncertainty on the $\mathcal{A}_{\rm CP}^{\rm dir, e}$ measurement in \eqref{eq:Adir_e}, this new relation holds great potential for providing a sharper picture of possible CP-violating NP.

In order to illustrate the potential difference between $R_K$ and $\bar{R}_{{K}}$, we consider the cyan star benchmark for $C_{9\mu}^{\rm NP}$ in Table~\ref{tab:benchmark} and \begin{equation}
    C_{9e}^{\rm NP} = (C_{9\mu}^{\rm NP})^* = 0.50 \abs{C_{9}^{\rm SM}} e^{-i 160^\circ} \ ,
\end{equation}
which in the $q^2$ bin of $[1.1,6.0] \; \si{\giga\eV^2}$ induces a direct CP asymmetry of $\mathcal{A}_{\rm CP}^{\rm dir, \mu} = -0.038 \pm 0.007$ for the muons and $\mathcal{A}_{\rm CP}^{\rm dir, e} = +0.038 \pm 0.007$ for the electrons. For this example, we have $\langle R_K \rangle = 1.001$ 
in agreement with the SM prediction. At the same time, we have
\begin{equation}
    R_K = 1.079, \quad \quad \bar{R}_K = 0.928, \quad \quad \mathcal{A}_{\rm CP}^{R_K} = -0.075 \ ,
\end{equation}
revealing LFU violation. This example thus clearly shows the potential of this new method.

\clearpage
\section{Conclusions}\label{sec:con}

In our analysis, we have explored the space left for electron--muon universality following from data for $B\to K \ell^+\ell^-$ decays in view of the recent measurement of the $\langle R_K \rangle$ ratio by the LHCb collaboration. We considered two scenarios which are characterised by having NP contributions either exclusively to the $C_{9\mu}^{\rm NP}$ coefficient or satisfying the relation $C_{9\mu}^{\rm NP}=-C_{10\mu}^{\rm NP}$. The corresponding Wilson
coefficients were determined in such a way to accommodate the measured differential $B\to K \mu^+\mu^-$ rates integrated over appropriate $q^2$ intervals that are found below their SM predictions at the $3.5 \, \sigma$ level. Moreover, we utilised experimental constraints on the direct CP asymmetry of the
charged $B^+\to K^+\mu^+\mu^-$ modes to determine ranges for the short-distance coefficients.

In our studies, we have applied a model following from experimental data to describe the $c\bar c$ resonance regions. In particular the direct CP asymmetries are sensitive to these effects, while the mixing-induced CP asymmetries as well as the differential decay rates are more robust. By the time the measurements of CP violation will become available, we should have a better description of the hadronic resonance regions.

The new LHCb result $\langle R_K \rangle=0.949 \pm 0.05$, which is consistent with the SM value of~1, may naively seem to suggest that NP effects should
exhibit electron--muon universality to a good approximation. However, as we have demonstrated, this is actually not the case should the NP contributions be associated with new sources of CP violation. In such a situation, which may well be realised in Nature, we can still have a significant violation of the electron--muon universality at the level of the
Wilson coefficients. 

These effects may be revealed through future measurements of CP violation in the charged $B^+\to K^+\ell^+\ell^-$ decays and analyses of direct and mixing-induced CP violation in neutral $B^0\to K_{\rm S}\ell^+\ell^-$ channels. We find that large differences between these CP asymmetries for the electronic ($\ell=e$) and muonic $(\ell=\mu$) modes are possible, thereby offering an exciting new portal for probing electron--muon universality. The strategy to map out electron--muon universality proposed in this paper can also be applied to other $B$ decays originating from
$\bar b \to \bar s \ell^+\ell^-$ quark-level processes, such as the $B~\to~f_0(980) \ell^+\ell^-$, $B~\to~K^\ast \ell^+\ell^-$ and $B_s^0~\to~\phi \ell^+\ell^-$ channels, which offer further observables to obtain a broader picture.

In view of the significant space for a violation of the electron--muon universality through new CP-violating couplings, experimental searches for differences between CP asymmetries of decays originating from $b \to s e^+ e^-$ and $b \to s \mu^+ \mu^-$ processes are strongly encouraged at the future high-precision frontier. These studies will play a key role for bringing tests of LFU to the next level, leaving a lot of space for possible surprises.

\section*{Acknowledgements}
This research has been supported by the Netherlands Organisation for Scientific Research (NWO). 

\clearpage
\bibliographystyle{JHEP} 
\bibliography{refs.bib}
\end{document}